\documentclass[10pt]{article}
\usepackage{amssymb}  
\usepackage{bm}
\usepackage{amsfonts}
\usepackage{amsmath}
\usepackage{authblk}
\usepackage{pifont}
\usepackage{pstricks}
\usepackage{graphicx}
\usepackage{amsmath}
\usepackage{authblk}
\usepackage{pstricks} 
\usepackage{pstricks-add}
\usepackage{framed}
\usepackage{bm}
\usepackage[T1]{fontenc}
\usepackage{pifont}
\usepackage{caption}
\usepackage{placeins}
\usepackage{pstricks}
\usepackage{pdfpages}
\usepackage{amsfonts}
\usepackage{graphicx}
\usepackage{placeins}
\usepackage{amsmath}
\usepackage{authblk}
\usepackage{pstricks}
\usepackage{pstricks-add}
\usepackage{framed}
\usepackage{bm}
\newtheorem{prop}{Proposition}[section] 
 
\newtheorem{rem}{Remark}[section]

\newtheorem{teo}{Teorema}[section]

\newtheorem{propr}{Property}[section]

\def\p{\mathop{P}\limits}

\def\t2dot{\mathop{\theta}\limits}
\def\x2dot{\mathop{x}\limits}
\def\x2dot{\mathop{x}\limits}

\title{Synchronization of a double pendulum with moving pivots: a study of the spectrum}
\author{F.~Talamucci}
\affil{DIMAI, Dipartimento di Matematica e Informatica ``Ulisse Dini'',\\
Universit\`a degli Studi di Firenze, Italy\footnote{
corresponding author e-mail: federico.talamucci@unifi.it}}
\date{}

\def\et2dot{\mathop{\eta}\limits}
\def\bet2dot{\mathop{\beta}\limits}
\def\t2dot{\mathop{\theta}\limits}
\def\s2dot{\mathop{\sigma}\limits}
\def\d2dot{\mathop{\delta}\limits}
\def\l2dot{\mathop{\lambda}\limits}
\def\ps2dot{\mathop{\cal E}\limits}
\def\tet2dot{\mathop{\theta}\limits}
\def\bfy2dot{\mathop{\bf y}\limits}
\def\bfq2dot{\mathop{\bf q}\limits}
\def\bbfq2dot{\mathop{\bar {\bf q}}\limits}
\def\w2{\mathop{W}\limits}
\def\xgrande2dot{\mathop{\bf X}\limits}
\def\p02dot{\mathop{P}\limits}
\def\a2dot{\mathop{A}\limits}

\def\csi2dot{\mathop{\xi}\limits}

\begin{document}
\bibliographystyle{plain}

\setcounter{equation}{0}
\setcounter{ese}{0}
\setcounter{eserc}{0}
\setcounter{teo}{0}               
\setcounter{corol}{0}
\setcounter{propr}{0}

\maketitle

\bibliographystyle{plain}

\setcounter{equation}{0}
\setcounter{ese}{0}
\setcounter{eserc}{0}
\setcounter{teo}{0}               
\setcounter{corol}{0}
\setcounter{propr}{0}

\maketitle

\vspace{.5truecm}

\noindent
{\bf Abstract}. The model we consider consists in a double pendulum set, where the pivot points are free to shift along a horizontal line.
Moreover, the two pendula are coupled by means of a spring whose extremities connect two points of each pendulum, at a fixed distance from the corresponding pivot.

\noindent
The mathematical model is first written encompassing a large class of setting for the device
(different sizes, different physical properties, ...). In order to carry on the problem of synchronization via analytical me\-thods, we focus on the circumstance of identical pendula: in that case, some classical theorems concerning the zeroes of polynomial equations are used in order to locate the eigenvalues governing the process, so that the possibility of synchronization of the device can be better understood. 

\vspace{.5truecm}


\vspace{.5truecm}


\section{The mathematical model}

\noindent
The interest on non-linear oscillations problems is ever more increasing, due to the large scale of applications in many fields. In particular, the question whether the system will attain a state of synchronization (in--phase or anti--phase) is one of the major achievements. 

\noindent
Here, we refer to a basic model similar to the ones studied in \cite{fra}, \cite{kum}, \cite{dil}, where Kuramoto coupled  pendula are studied both analitically and numerically.
The system consists in a couple of pendula, interacting by way of a spring connecting the rigid oscillating bars.
With respect to the cited models, we omit the action of the motor torque controlling one of the pendula, but we let the pivoting points to slide on a horizontal support acting as a non--smooth constraint. Moreover, the elastic interaction is here caused by a spring connecting the pendula not necessarily at the pivoting points.

\noindent
The first part of the paper makes an effort in formulating rigorously the mathematical model in the most general background of pendula different in the physical properties and central spring forcing in a non--simmetrical way. We also pay some attention to list the stable configurations of the system.

\noindent
At a later time, since we are mainly focused on an analytical study of the problem, we have to release some assuptions moving to the case of identical pendula: the mathematical advantage is certainly significant, since the system can be variables disentangled with respect to the variables related to in--phase (difference of the angles) and to anti--phase (sum of the angles) synchronization. 

\noindent
Via linearization of the system of equations at the stable equilibrium, we mainly investigate the possibility of locating the eigenvalues of the linearized problem: such an information can bring into being the way to control the synchronization phenomena, on the strength of the values of the parameters entering the process. 

\begin{figure*}[h!] 
%\centering 
\includegraphics[scale=.3]{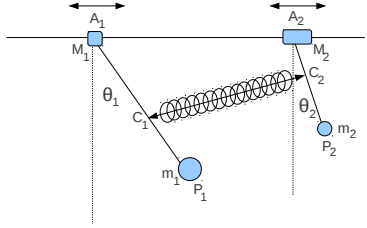} 
\caption{the experimental device} 
\label{modello}
\end{figure*}

\noindent
The physical model we are considering is outlined in Figure \ref{modello}, where the couple of pendula swinging on a vertical plane are sketched. The two bobs $P_1$ (mass $m_1$) and $P_2$ (mass $m_2$) oscillate with respect to the pivots $A_1$ (mass $M_1$) and $A_2$ (mass $M_2$), which can move on the horizontal line joining them. Setting the lenghts $\ell_i= \overline{A_iP_i}$, $i=1,2$, an interaction between the two pendula is exerted by a spring connecting $C_1\in A_1P_1$ with $C_2\in A_2P_2$, at the distances $\ell_{i,0}=\overline{A_iC_i}$, $i=1,2$. 
The latter quantities are assigned and fixed in each experiment.

\FloatBarrier

\noindent
The number of independent parameters which define the configuration of the system is clearly four: the two coordinates of $A_1$ and $A_2$ on the horizontal line and the two angles which $P_1$ and $P_2$ form with the vertical direction are feasible lagrangian coordinates. In that case, fixing a cartesian frame of reference such that the apparatus is contained in the vertical plane $y = 0$, the line $A_1A_2$ is the $x$--axis and the $z$--axis is upward--vertically directed and choosing ${\bf q}=(x_1, x_2, \theta_1, \theta_2)$ as the lagrangain parameters ($x_i$ coordinate of $A_i$, $\vartheta_i$ angle which $P_i-A_i$ forms with the downward vertical direction, $i=1,2$), the Lagrangian function of the sy\-stem is ${\cal L}({\bf q}, {\dot {\bf q}})=\dfrac{1}{2} {\dot {\bf q}}\cdot {\Bbb A}({\bf q}){\dot {\bf q}}-V({\bf q})$, 
where
\begin{equation}
\label{matra}
{\Bbb A}=\left( \begin{array}{cccc}
M_1+m_1 & 0 & m_1 \ell_1 \cos \theta_1 & 0 \\
0 & M_2+m_2 & 0 & m_2 \ell_2 \cos \theta_2 \\
m_1 \ell_1 \cos \theta_1 & 0 & m_1 \ell_1^2 & 0 \\
0 &  m_2 \ell_2 \cos \theta_2 & 0 & m_2 \ell_2^2
\end{array}
\right)
\end{equation}
and 
\begin{equation}
\label{pot}
V({\bf q})=-m_1g \ell_1\cos \theta_1 -m_2g\ell_2 \cos \theta_2 +\dfrac{1}{2} k\left(|C_1-C_2|-d\right)^2
\end{equation}
In (\ref{pot}) the lenght of the spring at rest is assumed to be $d\geq 0$. 

\noindent
Regarding the friction forces, whenever the damping is formulated as 
${\bm \Phi}_{A_i}=-\beta_i{\dot A}_i$, $i=1,2$
${\bm \Phi}_{P_i}=-\beta_{2+i}{\dot P}_i$, $i=1,2$, the lagrangian components of them are
\begin{equation}
\label{damping1}
{\bm \Phi}^{({\bf q})}=
\left(
\begin{array}{c}
-(\beta_1+\beta_3) {\dot x}_1 - \beta_3\ell_1 {\dot \theta}_1 \cos \theta_1\\
-(\beta_2+\beta_4) {\dot x}_2 -\beta_4 \ell_2 {\dot \theta}_2 \cos \theta_2\\
-\beta_3 \ell_1 {\dot x}_1 \cos \theta_1 - \beta_3 \ell_1^2 {\dot \theta}_1\\  
-\beta_4 \ell_2 {\dot x}_2 \cos \theta_2  -\beta_4 \ell_2^2{\dot \theta}_2
\end{array}
\right)=-
{\Bbb D}({\bf q}){\dot {\bf q}}
\end{equation}
where the elements of the positive definite matrix ${\Bbb D}$ are easily deduced. 
We assume that $B_i$, $i=1,2,3,4$ in (\ref{damping1}) are constant. 
Since ${\Bbb D}$ is also symmetric, we observe that the friction term can be expressed by means of a 
kynetic potential ${\cal D}({\bf q}, {\dot {\bf q}})=
\dfrac{1}{2}{\dot {\bf q}}\cdot {\Bbb D}{\dot {\bf q}}$ as
\begin{equation}
\label{potcin}
-{\Bbb D}({\bf q}){\dot {\bf q}} = -\nabla_{\dot {\bf q}}{\cal D}({\bf q}, {\dot {\bf q}}).
\end{equation}

\noindent
The equations of motion $\dfrac{d}{dt} (\nabla_{\dot {\bf q}}{\cal L})-\nabla_{\bf q}{\cal L}={\bm \Phi}^{({\bf q})}$ write explicitly
\begin{equation}
\label{em}
\left\{
\begin{array}{l}
\dfrac{d}{dt}\left( (M_1+m_1){\dot x}_1+m_1\ell_1 {\dot \theta}_1 \cos \theta_1\right)
+k(x_1-x_2+\ell_{1,0}\sin\theta_1 \\
- \ell_{2,0}\sin \theta_2){\cal E}({\bf q})=
-(\beta_1+\beta_3) {\dot x}_1 - \beta_3\ell_1 {\dot \theta}_1 \cos \theta_1, \\[3pt]
\dfrac{d}{dt}\left( (M_2+m_2){\dot x}_2+m_2\ell_2 {\dot \theta}_2 \cos \theta_2\right)
-k(x_1-x_2+\ell_{1,0}\sin\theta_1 \\
- \ell_{2,0}\sin \theta_2){\cal E}({\bf q})=
-(\beta_2+\beta_4) {\dot x}_2 -\beta_4 \ell_2 {\dot \theta}_2 \cos \theta_2, \\[3pt]
m_1\ell_1^2
{\t2dot^{..}}_1+m_1\ell_1{\x2dot^{..}}_1\cos \theta_1 +m_1g\ell_1 \sin \theta_1 +k\ell_{1,0}\left(
\ell_{2,0} \sin (\theta_1-\theta_2)\right.\\
\left.+(x_1-x_2)\cos \theta_1 \right){\cal E}({\bf q})=
-\beta_3 \ell_1 {\dot x}_1 \cos \theta_1 - \beta_3 \ell_1^2 {\dot \theta}_1, \\[3pt]
m_2\ell_2^2{\t2dot^{..}}_2+m_2\ell_2{\x2dot^{..}}_2\cos \theta_2 +m_2g\ell_2 \sin \theta_2 -k\ell_{2,0}\left(\ell_{1,0} \sin (\theta_1-\theta_2)\right.\\[3pt]
\left.+(x_1-x_2)\cos \theta_2 
\right){\cal E}({\bf q})=
-\beta_4 \ell_2 {\dot x}_2 \cos \theta_2  -\beta_4 \ell_2^2{\dot \theta}_2
\end{array}
\right.
\end{equation}
with 
\begin{equation}
\label{e}
{\cal E}({\bf q})= \dfrac{ |C_1-C_2| -d}{|C_1-C_2|}
\end{equation}
and $|C_1-C_2|=
[(x_1-x_2)^2+\ell_{1,0}^2+
\ell_{2,0}^2-2\ell_{1,0}\ell_{2,0} \cos (\theta_1 - \theta_2)+2(x_1-x_2)\times$ $(\ell_{1,0}\sin \theta_1 - \ell_{2,0} \sin \theta_2)]^{1/2}$.
The case $d=0$ corresponds to ${\cal E}\equiv 1$ in system (\ref{em}).

\section{Equilibrium}

\noindent
Having in mind to perform a linear approximation of the system (small oscillations), it is necessary to inspect the equilibrium positions and the possible stable feature. For this purpose, we start by remarking that, 
by summing up first and second equation in (\ref{em}), one gets
\begin{equation}
\label{eqxi}
M\dfrac{d^2 \xi}{dt^2}
=
-(\beta_1+\beta_3) {\dot x}_1-(\beta_2+\beta_4) {\dot x}_2
-
\beta_3\ell_1 {\dot \theta}_1 \cos \theta_1
-\beta_4 \ell_2 {\dot \theta}_2 \cos \theta_2
\end{equation}
where we defined 
\begin{equation}
\label{xi}
\xi=\dfrac{1}{M}\left[ (M_1+m_1)x_1+(M_2+m_2)x_2+m_1 \ell_1 \sin \theta_1 +m_2 \ell_2 \sin \theta_2\right] 
\end{equation}
which corresponds to the $x$--coordinate of the centre of mass of the system. 
On the other hand, by defining also 
\begin{equation}
\label{eta}
\eta=x_1-x_2
\end{equation}
the potential energy (\ref{pot}) is now written as (see also (\ref{e}))
\begin{eqnarray}
\label{poteta}
V=&-&m_1g \ell_1\cos \theta_1 -m_2g\ell_2 \cos \theta_2 +\dfrac{1}{2} k\{
[\eta^2+\ell_{1,0}^2+
\ell_{2,0}^2\\
&-&2\ell_{1,0}\ell_{2,0} \cos (\theta_1 - \theta_2)+2\eta(\ell_{1,0}\sin \theta_1 
- \ell_{2,0} \sin \theta_2)]^{1/2}-d\}^2
\nonumber
\end{eqnarray}
independently of $\xi$. The meaning of (\ref{eqxi}) and (\ref{poteta}) is evident: equilibrium makes sence with respect to the motion of $\xi$ (for instance, in absence of damping the velocity of the centre of mass is conserved) and the stationary positions of (\ref{poteta}) with respect to $(\eta, \theta_1, \theta_2)$ 
will produce the equilibrium configurations (relative to $\xi$) we are looking for. 

\noindent
Although we will focus on a specific stable configuration as the reference one for our analysis, it is worth to shortly examine the various possibilites of setting equilibrium: this will make clearer when the choosen reference   position is stable and render rigorous the approach by means of small oscillation. 

\noindent
The conditions for equilibrium $\nabla_{(\eta, \theta_1, \theta_2)}V={\bf 0}$ (see (\ref{poteta})) consists in 
\begin{equation}
\label{eq}
\begin{array}{l}
k(\eta+\ell_{1,0}\sin \theta_1 - \ell_{2,0}\sin \theta_2){\cal E}(\eta, \theta_1, \theta_2)=0,  \\
\alpha_1  +\left(\ell_{2,0} 
\sin (\theta_1 - \theta_2)+(x_1-x_2)\cos \theta_1
\right){\cal E}(\eta, \theta_1, \theta_2)=0, 
\\
\alpha_2-\left(\ell_{1,0} \sin 
(\theta_1 - \theta_2)+\eta\cos \theta_2\right){\cal E}(\eta, \theta_1, \theta_2)=0  
\end{array}
\end{equation}
where $\alpha_i =\dfrac{m_i g\ell_i}{k\ell_{i,0}}$, $i=1,2$ and   ${\cal E}(\eta, \theta_1, \theta_2)$ is achieved by means of (\ref{e}).

\begin{figure}[h!] 
%\centering 
\includegraphics[scale=.24]{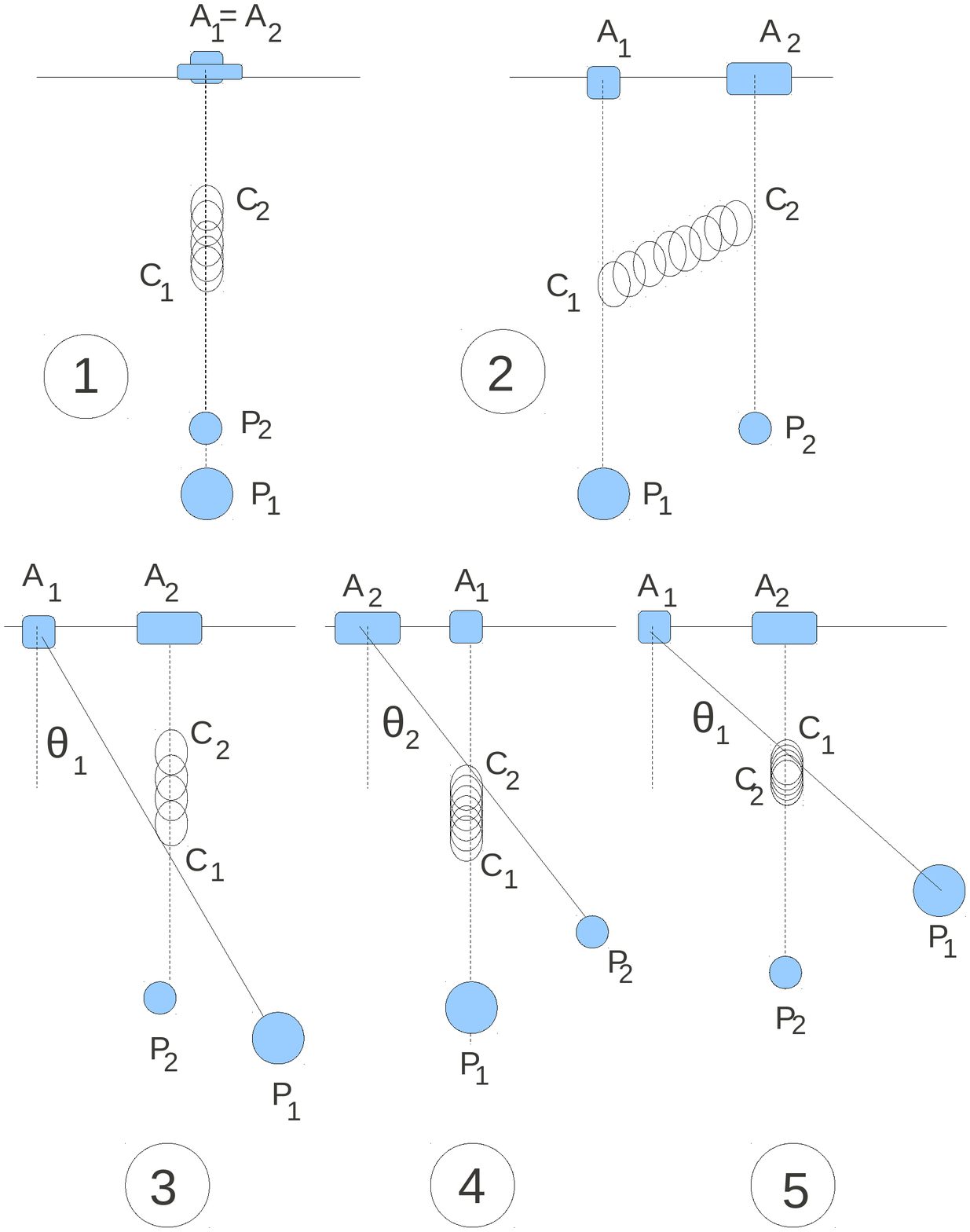} 
\caption[equil]{The equilibrium configurations are sorted by cases $\fbox{A}$: $\ell_{1,0}-\ell_{2,0}>d$ and $\fbox{B}$: 
$0\leq \ell_{1,0}-\ell_{2,0}\leq d$.

\noindent
Case \fbox{A} is consistent with \ding{172} ($C_2$ over $C_1$, stretched spring), \ding{174} ($C_2$ over $C_1$, stretched spring) and \ding{176} ($C_1$ over $C_2$, compressed spring).

\noindent
Case \fbox{B} is consistent with \ding{172} ($C_2$ over $C_1$, compressed spring), \ding{173} (only when $d>0$, spring at rest), \ding{175} 
for $\ell_{1,0} - \ell_{2,0} \leq d- \alpha_2  \leq \ell_{1,0}+\ell_{2,0}$, 
($C_2$ over $C_1$, compressed spring), \ding{176} 
for $0 \leq d- \alpha_1 \leq \ell_{1,0}+\ell_{2,0}$, 
($C_1$ over $C_2$, compressed spring).
}

\label{equilibrium}
\end{figure}

\noindent
Without loss of generality, 
let $\ell_{1,0}\geq \ell_{2,0}$ and consider only the solutions of (\ref{eq}) positioned in the lower half--plane $z\leq 0$: the equilibrium configurations where both pendula remain vertical are
$$
\begin{array}{lll}
\textrm{\ding{172}}\;
%[\textrm{from case}\;(ii)\textrm{--}a]\; 
&  \textrm{in any case:}& 
\eta=0, \; \;\;
\theta_1=\theta_2=0  \\
\textrm{\ding{173}}\;
%[\textrm{from case}\;(i)]\; 
& \textrm{if}\;\; \ell_{1,0}-\ell_{2,0}<d: &
|\eta|=\left( d^2-(\ell_{1,0}-\ell_{2,0})^2\right)^{1/2}, \;\;
\theta_1= \theta_2 =0. 
\end{array}
$$
There are other possibilities of equilibrium, which are listed hereafter and shown in Figure \ref{equilibrium}:
$$
\begin{array}{l}
\textrm{\ding{174}}\;\;
%[\textrm{from case}\;(ii)\textrm{--}c]
\quad 
\textrm{if}\;\;\; d+ \alpha_1\leq \ell_{1,0}-\ell_{2,0}: \\
\theta_2 =0, \;
\eta=-\ell_{1,0}\sin \theta_1, \;
\cos \theta_1 =\dfrac{1}{\ell_{1,0}}\left(
\ell_{2,0} + d+ \alpha_1 \right) \\
\textrm{\ding{175}}\;\;
%[\textrm{from case}\;(ii)\textrm{--}b] 
\quad 
\textrm{if}\;\;\; \ell_{1,0} - \ell_{2,0} \leq d- 
\alpha_2\leq \ell_{1,0}+\ell_{2,0}:\\
\theta_1 =0, \; \eta=\ell_{2,0}\sin \theta_2, \;
\cos \theta_2 = \dfrac{1}{\ell_{2,0}}
\left(
\ell_{1,0} - d+\alpha_2
\right)\\
\textrm{\ding{176}}\;\;
%[\textrm{from case}\;(ii)\textrm{--}c] 
\quad 
\textrm{if}\;\;\;-(\ell_{1,0}-\ell_{2,0}) \leq d-\alpha_1\leq \ell_{1,0}+\ell_{2,0}: \\
\theta_2 =0, \;
\eta=-\ell_{1,0}\sin \theta_1, \;
\cos \theta_1 =\dfrac{1}{\ell_{1,0}}\left(
\ell_{2,0} -d+\alpha_1\right) 
\end{array}
$$

\FloatBarrier

\noindent
Each of the positions from \ding{172} to \ding{174} have the symmetrical configuration, where $\eta$ is opposite in sign.

\noindent
Regarding stability, the analysis of the Hessian matrix $J_{\bf q}(\nabla_{\bf q}V)$ makes us conclude as follows:

\begin{itemize}
\item[$\bullet$] if $\ell_{1,0}-\ell_{2,0}<d$, then \ding{172} is unstable, position \ding{173} exists is stable, position \ding{174} does not exist, 
\item[$\bullet$] if $d<\ell_{1,0}-\ell_{2,0}<d+\alpha_1 $ then \ding{172} is stable and \ding{173}, \ding{174} do not exist,
\item[$\bullet$] if $\ell_{1,0}-\ell_{2,0}>d+\alpha_1$, then \ding{172} is unstable, position \ding{173} does not exist, position \ding {174} exists and the latter one is stable.
\end{itemize} 

\noindent
Owing to the effect of the damping forces, the stable positions are also asymptotically stable.

\noindent
Particular cases are $\ell_{1,0}-\ell_{2,0}=d$: then \ding{172} and \ding{173} overlap,
$d=0$: then  only \ding{172} and \ding{176} when $\alpha_1 \leq \ell_{1,0}-\ell_{2,0}$,
$\ell_{1,0}=\ell_{2,0}$: then \ding{176} is not possible, 
$d=0$, $\ell_{1,0}=\ell_{2,0}$: then equilibrium is only for \ding{172}.

%$$
%\begin{array}{l}
%\ell_{1,0}-\ell_{2,0}=d \;\; \Longrightarrow\;\;
%J_{\bf q}(\nabla_{\bf q}V)=\left(
%\begin{array}{cccc} 
%0 & 0 & 0 & 0 \\
%0 & 0 & 0 & 0 \\
%0 & 0 & m_1g\ell_1 & 0 \\
%0 & 0 & 0 & m_2g \ell_2
%\end{array}
%\right)
%\end{array}
%$$

\section{Synchronization: the equations of motion}

\noindent
In order to better face the problem of possible synchronization of the system, we make use of the variables 
\begin{equation}
\label{barq}
{\bar {\bf q}}={\bar {\bf q}}({\bf q}), \qquad {\bar {\bf q}}=(\xi, \eta, \sigma, \delta)
\end{equation}
where $\xi$ and $\eta$ are defined in (\ref{xi}), (\ref{eta}) and 
$$
\begin{array}{l}
\sigma = \theta_1 +\theta_2, \quad
\delta=\theta_1 - \theta_2
\end{array}
$$

\noindent
The onset of in--phase synchronization corresponds to solutions where to $\delta (t)\rightarrow 0$, on the other hand the anti--phase synchronization corresponds to $\sigma (t)\rightarrow 0$.

\noindent
We notice that transformation (\ref{barq}) can be inverted everywhere, since $det\,J_{\bf q}{\bar {\bf q}}=2$. In particular:
\begin{equation}
\label{thetainv}
\begin{array}{ll}
\theta_1 = \dfrac{\sigma+\delta}{2}, & \theta_2 = \dfrac{\sigma-\delta}{2}.
\end{array}
\end{equation}
With respect to the new variables ${\bar {\bf q}}$, the equations of motion are of the form
(they are obtained by multiplying (\ref{em}) by $J_{\bar {\bf q}}^T{\bf q}$)

\begin{equation}
\label{embarqmatr}
{\overline {\Bbb A}}(\sigma, \delta) \bbfq2dot^{..}+
{\bf W}(\sigma, \delta, {\dot \sigma}, {\dot \delta})+
\nabla_{\bar {\bf q}}{\overline V}(\eta, \sigma, \delta)=-{\overline {\Bbb D}}(\sigma, \delta){\dot {\bar {\bf q}}}
\end{equation}
where the matrices 
${\overline {\Bbb A}}$ and   
${\overline {\Bbb D}}$ are related to those appearing in (\ref{matra}) and (\ref{damping1}) by means of
${\overline {\Bbb A}}=(J_{\bar {\bf q}}{\bf q})^T{\Bbb A}(J_{\bar {\bf q}}{\bf q})$ and 
${\overline {\Bbb D}}=(J_{\bar {\bf q}}{\bf q})^T{\Bbb D}(J_{\bar {\bf q}}{\bf q})$;  moreover ${\bf W}$, containing the quadratic terms with respect to ${\dot \sigma}$ and ${\dot \delta}$, has its general form in ${\bf W}=\left[
J_{\bar {\bf q}} ({\overline {\Bbb A}}{\dot {\bar {\bf q}}})-\dfrac{1}{2} 
[J_{\bar {\bf q}}({\overline {\Bbb A}}{\dot {\bar {\bf q}}})]^T
\right]{\dot {\bar {\bf q}}}$. Lastly, ${\overline V}=V({\bf q}({\bar {\bf q}}))$ (see (\ref{poteta})).

\noindent
Concerning the explicit writing of system (\ref{embarqmatr}), if one defines 
 $\Psi_i (\sigma, \delta; C_1, C_2)$,  $C_1, C_2 \in {\Bbb R}$, $i=1,2$ as
\begin{equation}
\label{psi12}
\begin{array}{l}
\Psi_1 (\sigma, \delta; C_1, C_2)=C_1 \cos (\sigma/2) \cos (\delta/2)+C_2 
\sin  (\sigma/2) \sin  (\delta/2), \\[7pt]
\Psi_2 (\sigma, \delta; C_1, C_2)=C_1 \sin  (\sigma/2) \cos (\delta/2)+C_2 \cos  (\sigma/2) \sin (\delta/2)
\end{array}
\end{equation}
(from now on we omit the arguments $\sigma$ and $\delta$ for the sake of brevity),
he will find the following elements for the matrices and the vectors in (\ref{embarqmatr}):
$$
\begin{array}{l}
{\overline {\Bbb A}}_{1,1}=M, 
{\overline {\Bbb A}}_{1,2}=
{\overline {\Bbb A}}_{1,3}=
{\overline {\Bbb A}}_{1,4}=0, 
{\overline {\Bbb A}}_{2,2}=M_P, 
{\overline {\Bbb A}}_{2,3}=\Psi_1 (B_{m/M}^-, - B_{m/M}^+), \\
{\overline {\Bbb A}}_{2,4}=\Psi_1(B_{m/M}^+, -B_{m/M}^-), 
{\overline {\Bbb A}}_{3,3}=A_m^+ - \frac{1}{M}  \Psi_1^2(B_m^+, -B_m^-), \\
{\overline {\Bbb A}}_{3,4}=A_m^- -\frac{1}{M}\Psi_1 (B_m^+, -B_m^-)\Psi_1(B_m^-, -B_m^+), 
{\overline {\Bbb A}}_{4,4}=A_m^+-\frac{1}{M} \Psi_1^2(B_m^-, -B_m^+)
\end{array}
$$

$$
\begin{array}{l}
{\overline {\Bbb D}}_{1,1}=\beta, 
{\overline {\Bbb D}}_{1,2}=C_\beta^-, 
{\overline {\Bbb D}}_{1,3} = \Psi_1 (B_\beta^+, - B_\beta^-)-
\dfrac{\beta}{M}\Psi_1 (B_m^+, -B_m^-), \\
{\overline {\Bbb D}}_{1,4} = \Psi_1 (B_\beta^-, - B_\beta^+)
-\dfrac{\beta}{M}\Psi_1 (B_m^-, -B_m^+), 
{\overline {\Bbb D}}_{2,2}=C_\beta^+, \\
{\overline {\Bbb D}}_{2,3} =\Psi_1(B_{\beta/M}^-,-B_{\beta/M}^+)-\frac{C_\beta^-}{M}
\Psi_1 (B_m^+, -B_m^-), \\
{\overline {\Bbb D}}_{2,4}= 
\Psi_1(B_{\beta/M}^+, -B_{\beta/M}^-)-\frac{C_\beta^-}{M} \Psi_1 (B_m^-, -B_m^+), \\
{\overline {\Bbb D}}_{3,3}=
\frac{1}{M} \Psi_1(B_m^+, -B_m^-)\left( \frac{\beta}{M}\Psi_1 (B_m^+, -B_m^-)
-2\Psi_1 (B_\beta^+, - B_\beta^-)\right) + A_\beta^+ ,\\ 
{\overline {\Bbb D}}_{3,4}=
\frac{1}{M} \Psi_1(B_m^-, -B_m^+)\left( \frac{\beta}{M}\Psi_1 (B_m^+, -B_m^-)-\Psi_1 (B_\beta^+, 
-B_\beta^-)\right) \\
\qquad -\frac{1}{M} \Psi_1(B_m^+, -B_m^-)
\Psi_1 (B_\beta^-, - B_\beta^+)+ A_\beta^-, \\
{\overline {\Bbb D}}_{4,4}=
\frac{1}{M} \Psi_1(B_m^-, -B_m^+)\left( \frac{\beta}{M}\Psi_1 (B_m^-, -B_m^+)
-2\Psi_1 (B_\beta^-,-B_\beta^+) 
+ A_\beta^+\right), \\
{\bf W}_1=0, {\bf W}_2=
-\frac{1}{2} \psi_2
( B_{m/M}^-, B_{m/M}^+) ({\dot \sigma}^2+{\dot \delta}^2)-
\Psi_2( B_{m/M}^+, B_{m/M}^-){\dot \sigma}{\dot \delta}, \\[7pt]
{\bf W}_3=
\frac{1}{2M} \Psi_1(B_m^+, -B_m^-)
\left[ \Psi_2(B_m^+, B_m^-)({\dot \sigma}^2+{\dot \delta}^2)+2\Psi_2(B_m^-, B_m^+){\dot \sigma}{\dot \delta}\right], \\[7pt]
{\bf W}_4=
\frac{1}{2M} \Psi_1(B_m^-, -B_m^+)
\left[ \Psi_2(B_m^+, B_m^-)({\dot \sigma}^2+{\dot \delta}^2)+2\Psi_2(B_m^-, B_m^+)
{\dot \sigma}{\dot \delta}\right]
\\
\nabla_{\bar {\bf q}}{\overline V}_1=0, 
\nabla_{\bar {\bf q}}{\overline V}_2=
k{\cal E}[\eta+2\Psi_2(\ell_0^-, \ell_0^+)], \\[3pt]
\nabla_{\bar {\bf q}}{\overline V}_3=
g\Psi_2(B_m^+, B_m^-)+k {\cal E}\eta
\Psi_2(\ell_0^-,-\ell_0^+), \\[3pt]
\nabla_{\bar {\bf q}}{\overline V}_4=
g\Psi_2(B_m^-, B_m^+)+
k {\cal E}[\eta \Psi_1(\ell_0^+, -\ell_0^-)+\ell_{1,0}\ell_{2,0}\sin\delta]
\end{array}
$$

\noindent
where ${\cal E}(\eta, \sigma, \delta)$ is calculated by means of (\ref{e}) and (\ref{thetainv}). The constant quantities with superscript $+$, $-$ are
\begin{equation}
\label{pm}
\begin{array}{l}
\ell_0^{\pm}=\frac{1}{2} (\ell_{1,0}  \pm \ell_{2,0}), 
 \quad 
\beta=\sum\limits_{i=1}^4\beta_i, \quad M_P=\frac{1}{M}(M_1+m_1)(M_2+m_2)
\\ 
A_m^{\pm}=\frac{1}{4}(m_1\ell_1^2\pm m_2 \ell_2^2),
\quad 
A_\beta^{\pm}=\frac{1}{4} (\beta_3 \ell_1^2\pm \beta_4 \ell_2^2), \\
B_m^{\pm}=\frac{1}{2} (m_1 \ell_1 \pm m_2 \ell_2), \quad 
B_\beta^{\pm}=\frac{1}{2} (\beta_3 \ell_1 \pm \beta_4 \ell_2),\\
B_{m/M}^{\pm}=\frac{1}{2M} \left[(M_2+m_2)m_1\ell_1 \pm (M_1+m_1)m_2\ell_2\right], \\  
B_{\beta/M}^{\pm}=\frac{1}{2M} \left[(M_2+m_2)\ell_1\beta_3  \pm 
(M_1+m_1)\ell_2\beta_4\right], \\
C_\beta^+=\frac{1}{M^2}\left[
(\beta_1+\beta_3)(M_2+m_2)^2+(\beta_2+\beta_4)(M_1+m_1)^2\right],\\
C_\beta^-=\frac{1}{M}\left[
(\beta_1+\beta_3)(M_2+m_2)-(\beta_2+\beta_4)(M_1+m_1)\right].
\end{array}
\end{equation}

\noindent
An overview on (\ref{embarqmatr}) shows that, as far as the pendula are more and more similar (so that the  constant quantities in (\ref{pm}) with superscript $-$, except for $\ell_{0}^-$, tend to zero), the variable 
$\sigma$ is coupled stronger to $\xi$ than to $\eta$, while $\delta$ is coupled stronger to $\eta$. 

\section{The linear approximation}

\noindent
We will investigate the mathematical problem of small oscillations focussing on the equilibrium con\-fi\-gu\-ra\-tion \ding{173} (see Paragraph 1.1). Assuming $\xi(0)=0$, we make reference to the position  (see (\ref{thetainv})) $\sigma = \delta =0$ and $\eta_{eq}=[d^2-(\ell_{1,0}-\ell_{2,0})^2]^{1/2}$.
Such a choice has to be considered the most appropriate one, since the position is stable and the condition of existence $\ell_{1,0}-\ell_{2,0}<d$ moves in the direction of our next assumption (needed in order to facilitate calculations) $\ell_{1,0}\sim \ell_{2,0}$.

\noindent
The standard second order approximation of the Lagrangian function and the remark (\ref{potcin}) make us write the linearized problem as

\begin{equation}
\label{eqlin}
\begin{array}{l}
{\overline {\Bbb A}}_0 \bbfq2dot^{..}
+{\overline {\Bbb V}}_0({\bar {\bf q}}-{\bar {\bf q}}_0)=-{\overline {\Bbb D}}_0{\dot {\bar {\bf q}}}, \;\;
{\bar {\bf q}}_0=(0,\eta_{eq}, 0,0),\\[7pt]
{\overline {\Bbb A}}_0={\overline {\Bbb A}}(0,0), {\overline {\Bbb D}}_0
={\overline {\Bbb D}}(0,0), 
{\overline {\Bbb V}}_0=\left.J_{\bar {\bf q}}(\nabla_{\bar {\bf q}}
{\overline V})\right\vert_{(\eta, \sigma, \delta)=(\eta_{eq}, 0,0)}.
\end{array}
\end{equation}
Calling $\kappa= k\dfrac{\eta_{eq}^2}{d^2}$, $M_P = \dfrac{1}{M} (M_1+m_1)(M_2+m_2)$ and referring again to (\ref{pm}) for the constant quantities, the explicit calculation leads to the entries
$$
\begin{array}{l}
({\overline {\Bbb A}}_0)_{1,1}=M, 
({\overline {\Bbb A}}_0)_{1,2}=
({\overline {\Bbb A}}_0)_{1,3}=
({\overline {\Bbb A}}_0)_{1,4}=0,
({\overline {\Bbb A}}_0)_{2,2}=M_P, 
({\overline {\Bbb A}}_0)_{2,3}=B^-_{m/M}, \\ ({\overline {\Bbb A}}_0)_{2,4}=B^+_{m/M}, 
({\overline {\Bbb A}}_0)_{3,3}= A_m^+- \frac{1}{M} \left( B_m^+\right)^2, 
({\overline {\Bbb A}}_0)_{3,4}=A_m^- - \frac{1}{M} B_m^+B_m^- \\
({\overline {\Bbb A}}_0)_{4,4}=A_m^+- \frac{1}{M} \left( B_m^-\right)^2, \\[7pt]
({\overline {\Bbb V}}_0)_{1,1}=
({\overline {\Bbb V}}_0)_{1,2}=
({\overline {\Bbb V}}_0)_{1,3}=
({\overline {\Bbb V}}_0)_{1,4}=0, 
({\overline {\Bbb V}}_0)_{2,2}=\kappa, ({\overline {\Bbb V}}_0)_{2,3}=\kappa\ell_0^-, \\
({\overline {\Bbb V}}_0)_{2,4}=\kappa\ell_0 ^+, 
({\overline {\Bbb V}}_0)_{3,3}=\frac{1}{2} g B_m^+ + \kappa (\ell_0^-)^2, 
({\overline {\Bbb V}}_0)_{3,4}=\frac{1}{2} g B_m^- + \kappa \ell_0^+ \ell_0^-,\\
({\overline {\Bbb V}}_0)_{4,4}=
\frac{1}{2} g B_m^+ + \kappa (\ell_0^+)^2, \\
({\overline {\Bbb D}}_0)_{1,1} =\beta, 
({\overline {\Bbb D}}_0)_{1,2}=C_\beta^-, 
({\overline {\Bbb D}}_0)_{1,3}=B_\beta^+ - \frac{\beta}{M} B_m^+, 
({\overline {\Bbb D}}_0)_{1,4} = B_\beta^- - \frac{\beta}{M} B_m^- \\
({\overline {\Bbb D}}_0)_{2,2}=C_\beta^+, 
({\overline {\Bbb D}}_0)_{2,3}=
B_{\beta/M}^- - \frac{1}{M} C_\beta^- B_m^+, 
({\overline {\Bbb D}}_0)_{2,4}=
B_{\beta/M}^+ - \frac{1}{M} C_\beta^- B_m^-, \\
({\overline {\Bbb D}}_0)_{3,3}=
A_\beta^+ +\frac{1}{M} B_m^+ ( \frac{\beta}{M} B_m^+ - 2B_\beta^+), 
({\overline {\Bbb D}}_0)_{3,4}=
A_\beta^-
+\frac{1}{M}B_m^-( \frac{\beta}{M}B_m^+-B_\beta^+) \\
\qquad \qquad - \frac{1}{M}B_m^+B_\beta^-,
({\overline {\Bbb D}}_0)_{4,4}=
A_\beta^++\frac{1}{M} B_m^-( \frac{\beta}{M} B_m^- - 2 B_\beta^-).
\end{array}
$$

%\begin{rem} stessa cosa con approssimazione diretta su (\ref{psiapprox})?
%\end{rem}

\noindent
Since our interest is focussed on specific solutions of (\ref{eqlin}) (those producing synchronization), we need to put the system in an explicit form: the standard procedure performed by setting ${\bf y}=\left( \begin{array}{c}{\bar {\bf q}}-{\bar {\bf q}}_0 \\ {\dot {\bar {\bf q}}}\end{array}\right)$ yields 

\begin{equation}
\label{fn}
\begin{array}{ll}
{\dot {\bf y}}={\Bbb M}_0{\bf y}, & {\Bbb M}_0=\left( \begin{array}{cc} {\Bbb O} & {\Bbb I} \\ 
- {\overline {\Bbb A}}_0^{-1}{\overline {\Bbb V}}_0 & -{\overline {\Bbb A}}_0^{-1}{\overline {\Bbb D}}_0
\end{array}
\right)
\end{array}
\end{equation}
where ${\Bbb I}$, ${\Bbb O}$ are the null and the identity matrix of order $4$. As it is known,
the characteristic polynomial ${\cal P}(\lambda)$ connected to (\ref{fn}) is 
\begin{equation}
\label{polcar}
det\;\left( \lambda^2{\overline {\Bbb A}}_0+\lambda{\overline {\Bbb D}}_0+{\overline {\Bbb V}}\right)=0.
\end{equation}

\section{Identical pendula}

\noindent
When the physical characteristics of the two pendula are exaclty the same, namely
\begin{equation}
\label{pendoliuguali}
\begin{array}{lll}
m_1=m_2={\overline m}, & M_1=M_2={\overline M}, & \ell_1=\ell_2=\ell, \\
\beta_1=\beta_2={\overline \beta}, & \beta_3=\beta_4={\hat \beta}, &
\end{array}
\end{equation}
the  quantities in (\ref{pm}) with superscript $-$ (except for $\ell_{0}^-$) vanish and those with superscript $+$ are
\begin{equation}
\label{pmuguali}
\begin{array}{l}
%\ell_0^{\pm}=\dfrac{1}{2} (\ell_{1,0}  \pm \ell_{2,0}), \quad 
M=2({\overline M}+{\overline m}), \; 
\beta=2({\overline \beta}+{\hat \beta}), \; M_P=\frac{1}{2}({\overline M}+{\overline m}), \;
C_\beta^+=\frac{1}{2}({\overline \beta}+{\hat \beta}), \\
A_m^+=\frac{1}{2}{\overline m}\ell^2, \;
A_\beta^+=\frac{1}{2} {\hat \beta}\ell^2, \;
B_m^+={\overline m}\ell, \; 
B_\beta^+={\hat \beta} \ell,\;\\
B_{m/M}^+=\frac{1}{2}{\overline m}\ell, \; 
B_{\beta/M}^+=\frac{1}{2} {\hat \beta}\ell.
\end{array}
\end{equation}

\noindent
Under assumption (\ref{pendoliuguali}) of identical pendula
\begin{description}
\item[$\bullet$] $\eta$ is eliminated from the $\xi$--equation (first line of (\ref{embarqmatr})) and $\xi$ is eliminated from the $\eta$--equation (second line of (\ref{embarqmatr})),  
\item[$\bullet$] the simplifications in (\ref{psi12}) 
$$
\begin{array}{l}
\Psi_1 (\sigma, \delta; C_1^+, C_2^-)=C_1^+ \cos (\sigma/2) \cos (\delta/2), \\
\Psi_1 (\sigma, \delta; C_1^-, C_2^+)=C_2^+ \sin (\sigma/2) \sin (\delta/2), 
\\
\Psi_2 (\sigma, \delta; C_1^+, C_2^-)=C_1^+ \sin  (\sigma/2) \cos (\delta/2), \\
\Psi_2 (\sigma, \delta; C_1^-, C_2^+)=C_2^+ \cos  (\sigma/2) \sin (\delta/2),
\end{array}
$$
where $C_1^+$, $C_2^+$, $C_1^-$, $C_2^-$ are any constant in (\ref{pm}) with the same superscript, 
carry out the stronger or weaker couplings we pointed at.
\end{description}

\noindent
In order to study system (\ref{fn}), the simplification arising whenever assumption (\ref{pendoliuguali}) is assumed is remarkable: in that case, as the constant quantities (\ref{pm}) with superscript $-$ (except for $\ell_0^-$) vanish, the matrices in (\ref{embarqmatr}) reduce to special structures. 
 
\noindent
Indeed, in case of (\ref{pendoliuguali}), the entries $({\overline {\Bbb A}}_0)_{i,j}$ $({\overline {\Bbb D}}_0)_{i,j}$ with $i+j$ 
odd number are null: this makes, as 
far as only the kinetic energy ${\overline {\Bbb A}}_0$ and the damping forces ${\overline {\Bbb D}}_0$ 
are concerned, system (\ref{eqlin}) uncoupled, separating $(\xi, \sigma)$ from $(\eta,\delta)$.

\noindent
When considering also the active forces ${\overline {\Bbb V}}_0$, we remark that the coupling between the two pairs of variables $(\xi, \sigma)$ and $(\eta, \delta)$ is due only to $\ell_0^-\not =0$ (i.~e.~different distances where the spring is fixed): in order to make it clear, it is worthwhile to write again the equations of motion at this stage, thas it (\ref{pendoliuguali}), hence (\ref{pmuguali}), but $\ell_0^-\not =0$:

\begin{equation}
\label{s10}
\left\{
\begin{array}{l} 
M\csi2dot^{..}=-\beta {\dot \xi} 
-\ell( {\hat \beta} - \beta\frac{\overline m}{M}) {\dot \sigma},\\
({\overline M}+{\overline m})\et2dot^{..}+{\overline M}\ell\d2dot^{..}
+2\kappa \left[ (\eta-\eta_{eq})+\ell_0^- \sigma +\ell_0^+\delta \right]
= -({\overline \beta}+{\hat \beta}){\dot \eta}
-{\hat \beta}\ell{\dot \delta},\\
\dfrac{1}{2}{\overline m}\ell^2\dfrac{\overline m}{{\overline M}+{\overline m}} \s2dot^{..} 
+\dfrac{1}{2}g{\overline M}\ell\sigma
+\kappa \left[\ell_0^-(\eta-\eta_{eq})+(\ell_0^-)^2 \sigma +\ell_0^+\ell_0^-\delta \right]
=\\
-\ell \left({\hat \beta}-\frac{\beta}{M} {\overline M}\right)  {\dot \xi}-
-\ell^2\left[ \frac{\overline m}{M}\left( \beta\frac{\overline m}{M}-
2{\hat \beta}\right) + \frac{1}{2} {\hat \beta}\right] 
{\dot \sigma}, 
\\
{\overline m}\ell\et2dot^{..} + {\overline m}\ell^2\d2dot^{..}+g {\overline m}\ell\delta 
+2\kappa \left[\ell_0^+(\eta-\eta_{eq})+\ell_0^+\ell_0^-\sigma +  (\ell_0^+)^2\delta \right]
=\\
-{\hat \beta}\ell{\dot \eta}-{\hat \beta}\ell^2 {\dot \delta}
\end{array}
\right.
\end{equation}

\noindent
Writing (\ref{polcar}) for (\ref{s10}), straightforward calculations lead to
\begin{equation}
\label{polcaruguali}
{\cal P}(\lambda)=\lambda {\cal P}_3(\lambda){\cal P}_4(\lambda)+(\ell_0^-)^2\lambda 
(\beta+M\lambda){\cal P}_2(\lambda).
\end{equation}
We explain the structure of ${\cal P}$ in (\ref{polcar}): as expected, there is a null eigenvector $\lambda=0$, the one associated to $\xi$. Moreover, it is easy to check that ${\cal P}_j$, $j=2,3,4$ are the following  polynomials in $\lambda$ of degree $j$:

$$
\begin{array}{l}
{\cal P}_4=\dfrac{1}{4}\ell^2 {\overline m}{\overline M}\lambda^4+
\dfrac{1}{4}\ell^2 ({\overline m}{\overline \beta}+{\overline M}{\hat \beta})\lambda^3
+\frac{1}{2}[M_PB_m^+g+\frac{1}{2} \ell^2 {\hat \beta}{\overline \beta} 
+\kappa ({\overline m}(\ell-\ell_0^+)^2\\
\qquad +{\overline M}(\ell_0^+)^2)]\lambda^2
+\frac{1}{2}\left[B_m^+C_\beta^+g +\kappa
\left( {\hat \beta}\left( \ell_0^+-\ell\right)^2+{\overline \beta}(\ell_0^+)^2
\right)
\right]\lambda
+\dfrac{1}{2}\kappa B_m^+g,
\\[5pt]
{\cal P}_3 = \ell^2{\overline m}{\overline M}\lambda^3 +
{\overline m}{\hat \beta}\ell^2 \left(\frac {\overline \beta}{\hat \beta}+
\frac{\overline M}{\overline m}\right)\lambda^2+
[M(\dfrac{1}{2}B_m^+g+\kappa (\ell_0^-)^2\kappa)+
\ell^2{\hat \beta}{\overline \beta}]\lambda\\
\qquad +\beta(\frac{1}{2} B_m^+g + \kappa (\ell_0^-)^2),
\\[5pt]
{\cal P}_2=  \kappa^2
[(2\ell_0^+B_{m/M}^+ - A_m^+-(\ell_0^+)^2M_P)\lambda^2 + 
(2\ell_0^+ B_{\beta/M} - A_\beta^+ -(\ell_0^+)^2C_\beta^+)\lambda\\
\qquad - B_m^+g/2]
\end{array}
$$

\noindent
As expected, the coefficients of the terms with even exponents in ${\cal P}_4$, $\lambda{\cal P}_3$ 
and ${\cal P}_2$ depend on the measure quantities of the system (masses, lenghts) and on the active forces, on the other hand the terms with odd exponents depend on the friction and damping coefficients and are cancelled if these effects are neglected.

\noindent
At this point, it is important to reaffirm what we remarked just before writing system (\ref{s10}): if $\ell_0^-$ vanishes, then the motion is entirely uncoupled with respect to the pair of variables $(\xi, \sigma)$ on the one hand (first and third equations in (\ref{s10}))
and the pair $(\eta, \delta)$ on the other hand (second and fourth equations of (\ref{s10})).
Still for $\ell_0^-=0$, it is immediate to realize that the characteristic polynomial for the sub--system in $(\xi, \sigma)$ is $\lambda {\cal P}_3$, as well as the one for the sub--system in $(\eta, \delta)$ is ${\cal P}_4$: 
the factorization (\ref{polcaruguali}) confirms such a property, since the gap $\ell_0^-$ plays the role of a sort of perturbation of the symmetrical case $A_1C_1=A_2C_2$.

%\noindent
%{\bf Check:} recalling (\ref{pmuguali}), we have
%$$
%\begin{array}{ll}
%M_PA_m^+-(B_{m/M}^+)^2=  &
%C_\beta^+A_m^+ + M_PA_\beta^+-2B_{m/M}^+B_{\beta/M}^+=\\
%C_\beta^+A_\beta^+ - (B_{\beta/M}^+)^2=&
%M_P(\ell_0^+)^2+A_m^+-2\ell_0^+B_{m/M}^+= 
%\\
%C_\beta^+(\ell_0^+)^2+A_\beta^+-2\ell_0^+ B_{\beta/M}^+ = 
%&
%MA_m^+-(B_m^+)^2=\\
%\beta A_m^++MA_\beta^+-2B_m^+B_\beta^+ = &
%\beta A_\beta^+-(B_\beta^+)^2 =\qquad \square
%\end{array} 
%$$

\section{Localization of the eigenvalues}

\noindent
We examine now the solutions of ${\cal P}(\lambda)=0$ (see (\ref{polcar})) in the special case $\ell_0^-=0$: 
By virtue of the splitting (\ref{polcaruguali}), the roots of ${\cal P}_4(\lambda)=0$ determine the motion of $(\eta, \delta)$ and the roots of ${\cal P}_3(\lambda)=0$ the motion of $(\xi, \sigma)$. 

\noindent
Let us define the adimensional quantities
\begin{equation}
\label{adim}
X=\dfrac{\overline M}{\overline m}, \quad Y=\dfrac{\overline \beta}{\hat \beta}, \qquad
\gamma_1= \dfrac{g}{\ell}\left(\dfrac{\overline m}{\hat \beta}\right)^2, \quad
\gamma_2 = 2\dfrac{k\ell}{{\overline m}g}, \qquad \nu=\dfrac{\ell_0^+}{\ell}
\end{equation}
and the $time^{-1}$ constant $\mu = \dfrac{\hat \beta}{\overline m}$:
the two polynomials can be written in the form:  
$$
\begin{array}{l}
\frac{4}{ {\overline m}{\overline M}\ell^2}\times
{\cal P}_4(\lambda)=\lambda^4+\sum\limits_{i=0}^3 a_i\lambda^i=\lambda^4
+\mu \frac{X+Y}{X}\lambda^3
+\mu^2 \left(
\gamma_1\gamma_2\frac{\nu^2 X+(1-\nu)^2}{X}+\right.\\
\left.
\frac{\gamma_1 (1+X)+ Y}{X}\right)\lambda^2+
\mu^3\gamma_1 \frac{1+Y+\gamma_2(\nu^2Y+(1-\nu)^2)}{X}
\lambda+\mu^4
\gamma_1^2\gamma_2\frac{1}{X},\\
\frac{1}{{\overline m}{\overline M}\ell^2}\times
{\cal P}_3(\lambda)= \lambda^3+ \sum\limits_{i=0}^2b_i\lambda^i
=\lambda^3+ \mu \frac{X+Y}{X}\lambda^2+\mu^2\frac{\gamma_1 (1+X)+ Y}{X}
\lambda+\mu^3\gamma_1 \frac{1+Y}{X}
\end{array}
$$
where the values of the coefficients $a_0$, $a_1$, $a_2$, $a_3$ and $b_0$, $b_1$, $b_2$ are easily deduced.

\begin{rem}
The asymptotical stability of the position we examine entails that the real parts of the roots of (\ref{polcaruguali}) are negative: anyhow, it can be easily checked, at least in the simplified case $\ell_0^-=0$, that, by implementing the Routh--Hurwitz stability criterion (see \cite{gan}), conditions 
$$
b_2b_1>b_0, \qquad a_3a_2>a_1, \quad a_1a_2a_3>a_1^2+a_3^2a_0
$$
ensure that all the roots lie in the left half complex plane.
\end{rem}

\noindent
At this point, we employ the Enestr\"om--Kakeya Theorem, as stated in \cite{ene}, \cite{kak}

\begin{teo} {\rm (E--K Theorem)} Let $p_n(\lambda)=a_0 + a_1 \lambda + \dots + a_{n-1}\lambda^{n-1}+a_n\lambda^n$ a polynomial with 
$a_j>0$ for any $j=0,\dots, n$. Then, all the zeros of $p_n$ are contained in the annulus
of the complex $z$--plane $\rho_m\leq |z| \leq \rho_M$, where 
\begin{equation}
\label{annulus}
\rho_m=\min\left\{ \dfrac{a_0}{a_1}, \dfrac{a_1}{a_2}, \dots \dfrac{a_{n-1}}{a_n}\right\}, \;
\rho_M=\max\left\{ \dfrac{a_0}{a_1}, \dfrac{a_1}{a_2}, \dots \dfrac{a_{n-1}}{a_n}\right\}
\end{equation}
\end{teo}
We refer to \cite{ene, kak} for the proof.

\noindent
In order to apply the Theorem, the calculate the ratio of the coefficients

\begin{equation}
\label{ratio}
\begin{array}{l}
a_3=\mu\dfrac{X+Y}{X}, \;
\dfrac{a_2}{a_3}=\mu \dfrac{\gamma_1 (1+X)+Y+\gamma_1\gamma_2 \Lambda(X)}{X+Y}, \\[5pt]
\dfrac{a_1}{a_2}=\mu \gamma_1\dfrac{1+Y+\gamma_2\Lambda(Y)}{\gamma_1(1+X)+Y+\gamma_1\gamma_2 \Lambda(X)}, \;
\dfrac{a_0}{a_1}=\mu \dfrac{\gamma_1\gamma_2}{1+Y+\gamma_2\Lambda(Y)}, \\[9pt]
b_2=\mu\dfrac{X+Y}{X}, \;
\dfrac{b_1}{b_2}=\mu\dfrac{\gamma_1(1+X)+Y}{X+Y}, \;
\dfrac{b_0}{b_1}=\mu \gamma_1\dfrac{1+Y}{\gamma_1(1+X)+Y}.
\end{array}
\end{equation}
where $\Lambda$ is the linear function
$\Lambda(\zeta)=\nu^2\zeta+(1-\nu)^2$. 
According to our setting, we figure that in each experiment the spring stiffness, the lenght $\ell$, the mass ${\overline m}$, the coefficient ${\hat \beta}$ of the pendula and the placement $\ell_0^+$ are constant, so that $\gamma_1$, $\gamma_2$ and $\nu$ in (\ref{adim}) do not change. On the other hand, for the same set of data the properties ${\overline M}$ and ${\overline \beta}$ of the sliding pivots can be modified ($X$ and $Y$ in (\ref{adim})), in order to inspect the response of the apparatus.

\noindent  
Concerning the order of (\ref{ratio}), we prove the following
\begin{propr}
For any $X>0$, $Y>0$
\begin{equation}
\label{order}
\dfrac{b_1}{b_2}<\dfrac{a_2}{a_3}, \qquad
\dfrac{b_0}{b_1}<b_2,  \qquad \dfrac{a_1}{a_2}<a_3=b_2, \qquad \dfrac{a_0}{a_1}<\dfrac{a_2}{a_3}.
\end{equation}
\end{propr}
{\bf Proof}:
The first three inequalities in (\ref{order}) are immediate. The fourth one writes explicitly
$$
\gamma_1\nu_1^2XY+\nu_1Y^2
+\gamma_1(\nu_1\nu_2-\gamma_2)X+[\gamma_1(\nu_1\nu_2-\gamma_2)+\nu_2]Y+\gamma_1\nu_2^2>0
$$
where $\nu_1=1+\gamma_2\nu^2$, $\nu_2=1+\gamma_2(1-\nu)^2$.
It is not difficult to check that $\nu_1\nu_2\geq \gamma_2$ for any $\gamma_2>0$ and $\nu \in [0,1]$: this can be done, for instance, by remarking that $(\nu_1(\nu)\nu_2(\nu))^\prime =4\gamma_2 (\nu - 1/2)(\gamma_2\nu^2-\gamma_2 \nu +1)$ and realizing that $\nu_1(\nu)\nu_2(\nu)-\gamma_2$ is a nonnegative function. $\quad \square$

\noindent
By virtue of the previous property and calling ${\cal A}= 
\left\{ \dfrac{a_0}{a_1}, \dfrac{a_1}{a_2}, \dfrac{a_2}{a_3}, a_3\right\}$, 
${\cal B}=\left\{ \dfrac{b_0}{b_1}, \dfrac{b_1}{b_2},  b_2\right\}$, the test on the coefficients can be restricted to
$$
\begin{array}{l}
\min{\cal A}=
\min\left\{ \dfrac{a_0}{a_1}, \dfrac{a_1}{a_2}\right\}, \quad 
\max{\cal A}=
\max\left\{ \dfrac{a_2}{a_3}, a_3\right\}, \\[9pt]
\min{\cal B}=
\min\left\{ \dfrac{b_0}{b_1}, \dfrac{b_1}{b_2}\right\}, \quad 
\max{\cal B}=
\max\left\{ \dfrac{b_1}{b_2}, b_2\right\}
\end{array}
$$
By using the explicit expressions (\ref{ratio}), the information we need for ordering the coefficients corresponds 
to the following inequalities in $X$, $Y$:
$$
\begin{array}{l}
\dfrac{b_0}{b_1}<\dfrac{b_1}{b_2}\; \textrm{if and only if} \; \gamma_1^2X^2+\gamma_1XY+(1-\gamma_1)Y^2\\
\qquad \qquad +\gamma_1(2\gamma_1-1)X +\gamma_1 Y+\gamma_1^2>0\; \hfill \textrm{conic}\;\fbox{1} \\
\dfrac{b_1}{b_2}<b_2\; \textrm{if and only if} \; (1-\gamma_1)X^2+XY+Y^2-\gamma_1 X>0\;\hfill \textrm{conic}\;\fbox{2} 
\\[9pt]
\dfrac{a_0}{a_1}<\dfrac{a_1}{a_2} \; \textrm{if and only if} \; 
\nu_1^2Y^2-\gamma_1\gamma_2 \nu_1 X +(2\nu_1\nu_2-\gamma_2)Y\\
\qquad \qquad +\nu_2^2-\gamma_1\gamma_2 \nu_2>0 \; \hfill
\textrm{conic}\;\fbox{I} \\[9pt]
\dfrac{a_2}{a_3}<a_3 \; \textrm{if and only if} \; 
(1-\gamma_1\nu_1) X^2+XY+Y^2-\gamma_1 \nu_2 X>0 \; \hfill
\textrm{conic}\;\fbox{II}
\end{array}
$$
(we wrote ``conic $\fbox{k}\;$'', $k=1,2,I,II$, for indicating the curve defined by re\-pla\-cing $>$ with $=$ in the corresponding 
inequality). Let us examine hereafter the case $\gamma_1>1$ (see (\ref{adim})): this is a plausible case, for standard physical data.

\noindent
On the positive quarter ${\cal Q}=\{X>0, \;Y>0\}$ the two curves $\fbox{1}$ and $\fbox{2}$ are branches of different two conics (for $\gamma_1>1$ hyperboles) not intersecting each other.
The slope of the oblique asymptote of hyperbola $\fbox{1}$ [respectively $\fbox{2}\;$] is $m_1=\dfrac{\gamma_1}{\gamma_1-1} (\sqrt{\gamma_1-3/4}+1/2)$ [resp.~$m_2=\sqrt{\gamma_1-3/4}-1/2<m_1$]. The two branches have no intersection for any value of $\gamma_1$ in the whole quarter, as shown in Figure 2. Moreover, they keep the same profile independently of  $\gamma_2$, $\nu$. 

\begin{figure}[h!] 
%\centering 
\includegraphics[scale=.3]{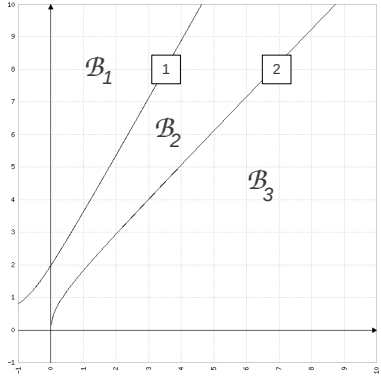} 
\caption[positive]{The positive quarter ${\cal Q}$ is splitted into three regions, according to the value of $max\,{\cal B}$ and $min\,{\cal B}$. Curves $\fbox{1}$ and $\fbox{2}$ do not match for any $\gamma_1>1$. The graphic is plotted with the value $\gamma_1=3.1$.} 
\label{1,2}
\end{figure}
\FloatBarrier

\noindent
Referring to the same figure:
$$
\begin{array}{lll}
\min{\cal B}=b_1/b_2, &  \max{\cal B}=b_2 &
\textrm{in}\;\;{\cal B}_1=\left\{(X,Y)\;\;\textrm{over}\;\;\fbox{1}\right\}
\\
\min{\cal B}=b_0/b_1, & \max{\cal B}=b_2 &
\textrm{in}\;\;{\cal B}_2=\left\{(X,Y)\;\;\textrm{under}\;\;\fbox{1}\;\;\textrm{and over}\;\;\fbox{2}\right\} 
\\
\max{\cal B}=b_0/b_1 & \min{\cal B}=b_1/b_2 &
\textrm{in}\;\;{\cal B}_3=\left\{(X,Y)\;\;\textrm{under}\;\fbox{2}\right\}
\end{array}
$$

\noindent
Analogously, conditions $\fbox{I}$ and $\fbox{II}$ split the quarter ${\cal Q}$ in regions showing different values for the minimum and the maximum of ${\cal A}$. 
The curve $\fbox{I}$ is an increasing branch of parabola whose axis of simmetry is parallel to the $X$--axis and vertex on the $X<0$ half--plane. Moreover, the intersection with the $X$--axis is positive [resp.~negative] according to $\nu_2 >\gamma_1\gamma_2$ [resp.~<], that is $\gamma_2<\left(\gamma_1-(1-\nu)^2\right)^{-1}$ [resp.~$>$].  
The curve $\fbox{II}$ is an increasing branch of a hyperbola passing through the origin with infinite slope. The asymptotic slope $m_{II}=\sqrt{\gamma_1\nu_1-3/4}-1/2\geq m_2$ ($=$ only for $\nu=0$) becomes greater than $m_1$ for large values of $\gamma_2$ and for $\nu\not =0$. Depending on the position of $(X,Y)\in {\cal Q}$, it is:
$$
\begin{array}{l}
\min{\cal A}=a_0/a_1, \, \max{\cal A}=a_3 \;
\textrm{for}\;\;{\cal A}_1=\left\{(X,Y)\;\;\textrm{over}\;\;\fbox{I}\;\;\textrm{and over}\;\;\fbox{II}\right\}
\\
\min{\cal A}=a_0/a_1, \, \max{\cal A}=a_2/a_3 \;
\textrm{for}\;{\cal A}_2=\left\{(X,Y)\;\textrm{over}\;\fbox{I}\;\textrm{and under}\;\fbox{II}\right\}
\\
\min{\cal A}=a_1/a_2, \, \max{\cal A}=a_3 \;
\textrm{for}\;{\cal A}_3=\left\{(X,Y)\;\textrm{under}\;\fbox{I}\;\textrm{and over}\;\fbox{II}\right\}
\\ 
\min{\cal A}=a_1/a_2, \, \max{\cal A}=a_2/a_3 \;
\textrm{for}\;{\cal A}_4=\left\{(X,Y)\;\textrm{under}\;\fbox{I}\;\textrm{and under}\;\fbox{II}\right\}
\end{array}
$$
whenever the regions exist.
Both curves $I$ and $II$ depend on the values of $\nu$, $\gamma_1$, $\gamma_2$ and the regions appear different according to them: basically, for $\nu_2\geq \gamma_1\gamma_2$, that is $\gamma_2\leq(\gamma_1-(1-\nu))^{-1}$, hyperbola $\fbox{I}$ remains under $\fbox{II}$, possibly with a local crossing via two intersections.
For $\gamma_2>(\gamma_1-(1-\nu))^{-1}$, either $\fbox{I}$ is over $\fbox{II}$ without any intersection, or 
$\fbox{II}$ passes over $\fbox{I}$, cutting it in one point of ${\cal Q}$ (this is the case plotted in Figure \ref{i,ii}).

\begin{figure}[h!] 
%\centering 
\includegraphics[scale=.3]{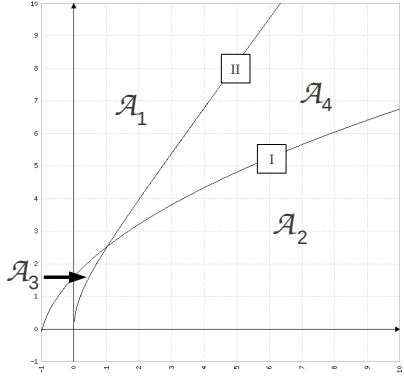} 
\caption{The regions on ${\cal Q}$ distincted by the values of $max\,{\cal A}$ and $min\,{\cal A}$. The curves are plotted by setting $\gamma_1=3.1$, $\gamma_2=2.2$, $\nu=0.4$.} 
\label{i,ii}
\end{figure}
\FloatBarrier

\noindent
The two pairs of curves $\fbox{1}$, $\fbox{2}$ and $\fbox{I}$, $\fbox{II}$ have to be traced together in order to compare the minima and the maxima. 

\noindent
Concerning the maximal radius, by (\ref{order}) it is $\dfrac{b_1}{b_2}<\dfrac{a_2}{a_3}$ and $b_2=a_3$. 
Moreover, it can be easily seen that in ${\cal Q}$ hyperbola $\fbox{II}$ is over $\fbox{2}$ for any $\nu\in[0,1]$ and $\gamma_2>0$ and they meet only in the origin: in other words, the region in ${\cal Q}$ where $a_2/a_3<a_3$ is contained in the region where $b_1/b_2<b_2$.
Thus, simply by overlapping the profiles of $\fbox{2}$ and $\fbox{II}$, the comparison between $\max{\cal A}$ and $\max{\cal B}$ is easily resolved in each point of ${\cal Q}$ as follows:

$$
\begin{array}{ll}
\max{\cal B}=b_2=max{\cal A}=a_3 &
\textrm{for}\;\;(X,Y)\;\;\textrm{over}\;\;\fbox{II}
\\
\max{\cal B}=b_2< \max{\cal A}=a_2/a_3 &
\textrm{for}\;\;(X,Y)\;\;\textrm{over}\;\;\fbox{2}\;\;\textrm{and under}\;\;\fbox{II}
\\
\max{\cal B}=b_1/b_2< \max{\cal A}=a_2/a_3 &
\textrm{for}\;\;(X,Y)\;\;\textrm{under}\;\;\fbox{2} 
\end{array}
$$

\begin{figure}[h!] 
%\centering 
\includegraphics[scale=.35]{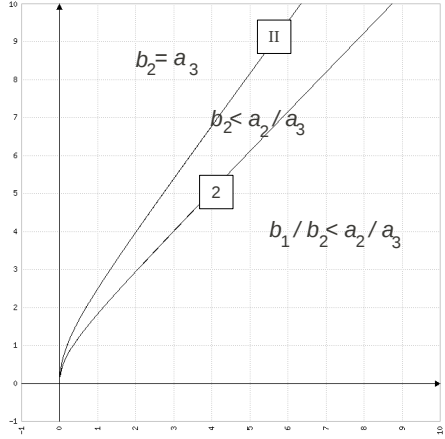} 
\caption[max]{comparison of the maximal radii all over ${\cal Q}$. The value of $max\,{\cal A}$ [resp.~$max\,{\cal B}$] changes by crossing the curve $\fbox{II}$ [resp.~$\fbox{2}$]. The two curves match only in the origin, whatever the quantities $\gamma_1$, $\gamma_2$, $nu$ are. In each of the three regions the value of $max\,{\cal A}$ and $max\,{\cal B}$ are compared. } 
\label{2,ii}
\end{figure}
\FloatBarrier

\noindent
The previous scheme together with Property 1.2, first inequality, entail the following 

\begin{propr}
The circle with radius $max{\cal B}$ containing the eigenvalues ${\cal P}_3=0$ (motion of $(\xi, \sigma)$) is in any case contained in the circle with radius $max{\cal A}$ containing the eigenvalues ${\cal P}_4=0$ (motion of $(\eta, \delta)$). The regions which locate the eigenvalues according to the E.~K.~theorem have a non-empty intersection, since $\dfrac{a_1}{a_2}<b_2$ (see (\ref{order})). 
\end{propr}

\begin{rem}
In other words, it is not possible to separate 
$(\min{\cal A}, \max{\cal A})$ (containing the eignevalues of ${\cal P}_4$) and $(\min{\cal B}, \max{\cal B})$ (containing the eigenvalues of ${\cal P}_3$), as it was in the model discussed in \cite{ft} of two oscillating pendula with pivots on a mobile support.
\end{rem}

\noindent
Since we are aiming to predict conditions which lead the system to a sharp tendency (in--phase or antiphase), we are induced to detect deeper where the eigenvalues are located, by enclosing them in narrower regions.
First of all, we check the following statement concerning the kind of the roots.

\begin{propr}
Set
\begin{equation}
\label{c12}
c_1=b_2-3\dfrac{b_1}{b_2}, \qquad c_2=b_2-9\dfrac{b_0}{b_1}.
\end{equation}
\begin{itemize}
\item[$(1)$] If either $c_1<0$ or $c_2<0$, then ${\cal P}_3$ has only one real root; this occurs in particular for $(X,Y)\in {\cal B}_3$.  
\item[$(2)$] ${\cal P}_3$ has three real roots in and only if $\alpha>0$, $\beta>0$ and 
$\left| \dfrac{3b_1}{2b_2} \dfrac{c_2}{c_1} -b_2\right| < \sqrt{b_2c_1}$.
\item[$(3)$] If either $\alpha =a_3- \dfrac{8a_2}{3a_3}<0$ or $\gamma=a_3-\dfrac{16a_0}{a_1}<0$ then ${\cal P}_4$ has at most two real roots; especially this occurs for $(X,Y)\in {\cal A}_2$ or  ${\cal A}_4$.
\end{itemize}
\end{propr}

\noindent
{\bf Proof}: it results immediately from the calculation of the Sturm sequence for the two polynomials: as
for ${\cal P}_3$ one finds $f_0={\cal P}_3$, $f_1=3\lambda^2+2b_2\lambda+\lambda_0$, $f_2=b_2\left(\dfrac{2}{3}c_1\lambda+\dfrac{b_1}{b_2}c_2\right)$, $f_3=-b_1+3b_1\dfrac{c_1}{c_2}-\dfrac{27}{4} \left(\dfrac{b_1}{b_2}\dfrac{c_2}{c_1}\right)^2$ and the evaluation at $\lambda=0$ and $\lambda\rightarrow -\infty$ gives $(1)$ and $(2)$. The same procedure for ${\cal P}_4$, up to the first three polynomial of the Sturm sequence, 
leads to $(3)$. $\quad \square$

\begin{rem}
Case $(2)$ in the previous Property and analogous conditions for ${\cal P}_4$ in order to have all real roots 
can be detected deeper by plotting on ${\cal Q}$ the regions where the roots of both ${\cal P}_3$ and ${\cal P}_4$ are negative real numbers: this will show a region alongside the positive $Y$--semiaxis, which can be  
described by $Y>>X$. Such a condition is plausible if one goes back to the physical quantities (\ref{adim}).
\end{rem}

\section{Controlling the eigenvalues of ${\cal P}_3$}

\noindent
As we already remarked, the localization based on the method (\ref{annulus}) does not exhibit a spontaneous separation of the roots of the two polynomial. Our final task consists in inspecting the possibility of enclosing the spectrum of one of the two polynomial in a specific region depending on the values of controllable parameters.

\noindent
Starting from the analysis developed above, we focus on  the region ${\cal B}_3$, where ${\cal P}_3$ definitely has one real root and two complex conjugate roots, since $c_1<0$. 

\noindent
We state the following
\begin{prop}
If $(X,Y)\in {\cal B}_3$ and $c_2>0$ (see (\ref{c12})), then 
the real parts of the roots $\lambda$ of ${\cal P}_3(\lambda)=0$ satisfy 
\begin{equation}
\label{re}
\Re(\lambda)<-\dfrac{1}{3}\dfrac{b_0}{b_1}.
\end{equation}
\end{prop}

\noindent
{\bf Proof}. Let us consider the derived point $\zeta$ of $\lambda=0$ for the polynomial ${\cal P}_3$:
$\zeta=0-3\dfrac{{\cal P}_3(0)}{{\cal P}_3^\prime (0)}=-3\dfrac{b_0}{b_1}$. As it is known, 
a Theorem of Laguerre (\cite{lag}) states that any circle passing through $\lambda=0$ and $\zeta$ is such that at least one root of ${\cal P}_3=0$ is internal to the circle, at least one root is external to it.

\noindent
Since $(X,Y)\in {\cal B}_3$, the coefficients are in the order 
$-\dfrac{b_1}{b_2}<-b_2<-\dfrac{b_0}{b_1}$. Moreover, $c_2>0$ implies $-3\dfrac{b_0}{b_1}>-b_2$. 
Now, we calculate the Sturm sequence at the values $\lambda=-b_2$ and $\lambda=-3\dfrac{b_0}{b_1}$:

\begin{center}
\begin{tabular}{c||c|c} 
 & $-b_2$ & $-3\dfrac{b_0}{b_1}$ \\[7pt] \hline \hline 
$f_0={\cal P}_3$ & $b_1(\frac{b_0}{b_1}-b_2)$ & $2b_0+3(\frac{b_0}{b_1})^2(\frac{1}{3}b_2-\frac{b_0}{b_1})$  \\[5pt] \hline
$f_1={\cal P}^\prime_3$ &  $b_2^2+b_1$ & $27(\frac{b_0}{b_1})^2-6b_2\frac{b_0}{b_1}+b_1$\\[5pt] \hline
$f_2$ & $b_2(-\frac{2}{3} b_2c_1 + \frac{b_1}{b_2}c_2)$ & 
$b_2(-2 \frac{b_0}{b_1}c_1 + \frac{b_1}{b_2}c_2)$  \\[5pt]  \hline
$f_3$  & $-b_1+3b_1\frac{c_1}{c_2}-\frac{27}{4} \left(\frac{c_2}{c_1}\right)^2(\frac{b_1}{b_2})^2$ & $-b_1+3b_1\frac{c_1}{c_2}-\frac{27}{4} (\frac{c_2}{c_1})^2(\frac{b_1}{b_2})^2$
\end{tabular}
\end{center}
Owing to the assumptions, for $\lambda=-b_2$ the signs of the sequence are $<0$, $>0$, $>0$, $<0$ hence two variations. On the other hand, since $c_1<0$ it is easy to check that $f_1>0$ at $\lambda=-3\dfrac{b_0}{b_1}$. Thus, the sequence of signs for $\lambda=-3\dfrac{b_0}{b_1}$ is $>0$, $>0$, $>0$, $<0$, giving one variation: we can conclude that the only real root of ${\cal P}_3=0$ lies between $-b_2$ and $-3\dfrac{b_0}{b_1}$. Lastly, if one consider the circle ${\cal C}$ with diameter $3\dfrac{b_0}{b_1}$ and centre in $\left(-\dfrac{3}{2}\dfrac{b_0}{b_1}, 0\right)$, the real root is external with respect to ${\cal C}$, so that the conjugate complex roots must be enclosed by ${\cal C}$. At the same time, the couple of roots is contained in the annulus with minimum radius $\rho_m=\dfrac{b_0}{b_1}$: since ${\cal C}$ intersects the internal boundary of the annulus at $X_i=- \dfrac{1}{3}\dfrac{b_0}{b_1}$, the module of the real part of the complex roots cannot place under $|X_i|$ (see Figure ), hence (\ref{re}) is proved. $\quad\square$

\noindent
The two regions where $c_1<0$ (encompassing ${\cal B}_3$) and where $c_2>0$ correspond respectively to 
$(1-3\gamma_1)X^2-XY+Y^2 - 3\gamma_1 X>0$, $\gamma_1 X^2+(1-8\gamma_1)XY+Y^2-8\gamma_1 X+\gamma_1 Y>0$. The two conics delimiting the regions in ${\cal Q}$ are $c_1=0$, which is a branch of hyperbola very similar to $\fbox{2}$ and standing above it with asymptotic direction $m_{c_1}= \sqrt{3\gamma_1-3/4}-1/2>m_2$, and $c_2=0$, two branches of a hyperbola with asymptotic slopes $m_{c_2^\pm}=[8\gamma_1 -1\pm \sqrt{(8\gamma_1-1)^2-4\gamma_1}]/2$.  For any set of values of the parameters it is $m_{c_2^-}<m_{c_1}<m_{c_2^+}$; moreover, $m_2>m_{c_2^-}$ only for $\gamma_1$ very close to $1$. In any case, the region on ${\cal Q}$ where $c_1<0$ and $c_2>0$, which is sketched in Figure \ref{c1,c2}.
is not empty for any assignment of $\gamma_1>1$, $\gamma_2$ and $\nu\in [0,1]$.

\begin{figure}[h!] 
%\centering 
\includegraphics[scale=.3]{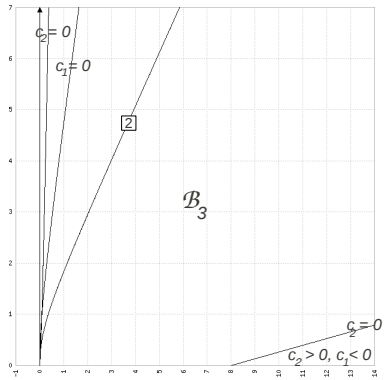} 
\caption[max]{ the region where $c_1<0$ is on the right w.~r.~t.~the curve marked by $c_1=0$ and encompasses ${\cal B}_3$, the region where $c_2>0$ is external w.~r.~t.~the curves marked with $c_2^\pm=0$. Apart from a narrow strip near the origin (actually the slope of $c_1=0$ is infinte at that point and the slope of $c_2^+$ is finite), the main part where $c_1<0\wedge c_2>0$ is on the right w.~r.~t.~$c_2^-=0$ and expands to infinite, by increasing $X$ and $Y$. } 
\label{c1,c2}
\end{figure}
\FloatBarrier

\noindent
The estimation (\ref{re}) provides a tool in order to induce a preponderant anti--phase disposition of the system: actually, by operating on the parameters entering $b_0/b_1$ it is possible to move away from $0$ the eigenvalues of ${\cal P}_3$, so that $\sigma$ rapidly diminishes. 

\noindent
On the other hand, the spectrum of ${\cal P}_4$, although it has not been investigated in depth, is independent on the one of ${\cal P}_3$, because of the presence of $\gamma_2$, $\nu_1$ and $\nu_2$, 
so that a way to sharply separate the two set of roots is traced.

\section{Conclusions}

\noindent
Our first purpose was to formulate accurately, via the Lagrangian formalism, the problem of coupled oscillation of two different pendula,  whose pivots are placed on masses moving along a track. The system is susceptible to sliding friction and to air resistance. The system of equations, whenever the features of the model are allowed to be general, is (\ref{em}), then (\ref{embarqmatr}), 
as soon as the change of variables (\ref{barq}) has been adopted. The corresponing mathematical problem, even though approximated as in (\ref{eqlin}), is anyway difficult to face from the analytical point of view, if the 
investigation concerns the settlement of in--phase or anti--phase synchronization.

\noindent
In writing explicitly the matrices of (\ref{eqlin}) we intended to highlight the coupling of the significant variables $\sigma$ and $\delta$ with the other ones: it appears clearly that the assumption of identical properties of the teo pendula gives rise to a drastic simplification, due essentially to the factorization of the characteristic polynomial. The approach we elaborated goes with the possibility in the experimental device
of modifying the masses sliding on the track and the friction on it.

\noindent
By employing classical theorens on the placement of complex roots of polynomials (Enestr\"om--Kakeya and Laguerre theorems), it is possible to trace two regions on the complex plane where the eigenvalues connected to $\sigma$ and $\delta$ are positioned.

\noindent
Contrary to the case analyzed in \cite{ft}, the two regions are not spontaneously separated and some operation to control them (an istance is (\ref{c12})) have to be performed in order to drive the system to synchronization.
The corresponding conditions can be explained in terms of the physical parameters.

\noindent
By preserving the same approach, other kind of responses of the system can be checked, as, for istance, the role of the distance of the extremity of the spring from the pivot point, modifying $\nu$.

\noindent
From the mathematical point of view, the application of some refined versions of the E--K theorem, 
as for istance the one in \cite{gel}, can help for a more restricted localization on the quarter ${\cal Q}$.

\noindent
On the other hand, the analytical study of the more complex case of different pendula can be performed via 
implementing a perturbation of the present case, by defining the deviations $\epsilon=G^+-G^-$ of any 
quantity $G^\pm$, and writing (\ref{embarqmatr}) with additional terms depending on them.

\noindent
Finally, we remark that from the numerical point of view, the method can provide a valid starting point in order to compute the spectrum and make the computer information more accurate: indeed, the selection of the initial data on the regions of ${\cal Q}$


\begin{thebibliography}{10}

\bibitem{dil} Dil\~ao, R.~,  On the problem of synchronization of identical dynamical systems:
The Huygens clocks. In
A.~Frediani and G.~Butazzo (Eds.), ''Variational
Methods in Aerospace Engineering'',
Springer Optimization and its Applications, Vol.~33, Cap.~10, 163
--181, Springer Verlag, 2009


%
\bibitem{ene}
Enestr\"om, G.~, Remarque sur un th\'eor\`eme relatif aux racines de l'equation
$a_nx^n+a_{n−1}x^{n−1}+\dots a_1 x+a_0=0$ o\`u tous les coefficientes
$a$ sont r\'eels et positifs, T\^ohoku Mathematical Journal {\bf 18}, 34--36, 1920

%
\bibitem{fra} Fradkov, A.~L.~, Andrievsky, B.~, Synchronization and phase relations in the motion of two--pendulums system, Int.~J.~of Non--Linear Mechanics, {\bf 42}, 895--901, 2007


\bibitem{gel} Anderson N.~, Saff E.~B.~, Varga R.~S.~, An extension of the Enestr\"om--Kakeya theorem and its sharpness, Siam J.~Math.~Anal.~, V.~12 N.~1, 10--22, 1981

%
\bibitem{kak}
Kakeya, S.~, On the Limits of the Roots of an Algebraic Equation with
Positive Coefficients, T\^ohoku Mathematical Journal (First Series) {\bf 2}, 140--142, 1912--13

%
\bibitem{kum} Kumon, M.~, Washizaki, R.~, Sato, J.~, Kohzawa, R.~, Mizumoto, I.~, Iwai, Z.~, 
Controlled synchronization of two $1$--DOF coupled oscillators, Proceedings of the $15$--th Triennal World Congress of IFAC, Barcelona, Spain 2002


\bibitem{gan} Gantmacher, F.~R.~, Applications of the theory of matrices,  translated and revised by J.~L.~Brenner, Interscience Publishers, Inc.~, New York, 1959


\bibitem{lag} Laguerre, E.~, Sur al r\'esolution des \'equations num\'eriques, Nouvelles Annales de math\'ematique, II, 17. In Oeuvres, Vol.~I, 53--63, 1878


\bibitem{ft} Talamucci, F.~, Synchronization of two coupled pendula in absence of e\-sca\-pe\-ment, Applied Mathematics and Mechanics (English Edition), Vol.~37, N.~12, 1721--1738, 2016


\end{thebibliography}
\end{document}